# Observational properties of extreme supernovae

C. Inserra*

School of Physics & Astronomy, Cardiff University, Cardiff, UK

**The last ten years have opened up a new parameter space in time-domain astronomy with the discovery of transients defying our understanding of how stars explode. These extremes of the transient paradigm represent the brightest - called superluminous supernova - and the fastest - known as fast, blue optical transients - of the transient zoo. The number of their discoveries and information gained per event have witnessed an exponential growth that has benefited observational and theoretical studies. The collected dataset and the understanding of such events have surpassed any initial expectation and opened up a future exploding with potential, spanning from novel tools of high-redshift cosmological investigation to new insights into the final stages of massive stars. Here, the observational properties of extreme supernovae are reviewed and put in the context of their physics, possible progenitor scenarios and explosion mechanisms.**

Portraying the landscape of extreme transients is not a trivial task. Almost a decade ago astronomers discovered transients defying the standard paradigm of stellar explosion. Their luminosity and evolution cannot be explained by the two classical mechanisms of core-collapse[1] and thermonuclear[2] explosions. Their observables, altogether, cannot be overall explained by the interaction between an ejected expanding medium (i.e. ejecta) and a circum-stellar material (CSM) previously expelled by the dying star (e.g. IIn/Ibn supernovae), nor by what is expected and observed in Tidal Disruption Events (TDEs), where a star is gravitationally disrupted by a black hole. In the transient parameter space of peak luminosity versus rise-time (Fig. 1) extreme transients lie above the lines representing the maximum possible luminosity from a standard supernova (SN) explosion. Such limits are determined using the standard diffusion formalism[3] and the maximum ratio of $^{56}$Ni mass with respect to that of the ejecta as derived from theoretical[4] and observational studies (average value)[5]. For such reasons, we refer to them as the extremes of the supernova (and transient) population. These objects can be grouped into two categories: 1 – a population of ultra-bright 'superluminous' supernovae (SLSNe), some 100 times brighter than classical supernova types, offering new probes of the high redshift universe and the potential for a new class of standard candle[6, 7] ; 2 – transients showing a fast rise and subsequent rapid decay (FBOTs or rapidly evolving transients) that do not resemble any common class of extragalactic transient[8, 9]. In this Review Article I review their observational properties showing why they are interesting and eventually focus on their future prospects. Note that evolutionary stages (phases) are with respect to maximum light and in the rest frame unless otherwise stated.

## 1 - Superluminous supernovae

It is usually reported that superluminous supernovae are characterized by absolute luminosities at maximum light of $M_{AB} \lesssim -21$ mag, total radiated energies of the order of $10^{51}$ erg[10] and a preference for low-metallicity, star-forming environments[11, 12, 13, 14, 15, 16]. There are two broad classes: SLSNe II, which exhibit signatures of hydrogen in their optical spectra, and SLSNe I, which do not. The former are heterogeneous in both luminosity and the host environment[12, 15, 17], and the majority of the population consists of events displaying signatures of interaction similar to classical SNe IIn (e.g., SN2006gy)[18], with a smaller contribution from intrinsically bright events (e.g., SNe 2008es, 2013hx)[19, 20, 21]. Hydrogen-poor SLSNe[6, 17, 22] are intrinsically rare, $10^{-4}$ of the rate of core-collapse SNe up to z~1[23, 24, 25] (although that might increase following the star formation history at higher redshift[23, 26]), and spectroscopically linked to normal or broad-line SNe Ic[27]. Their characteristic spectroscopic evolution and connection with massive star explosions have been the most distinctive attribute of SLSNe I, together with their typical explosion location in dwarf, metal-poor, and star-forming galaxies[11, 12, 13, 14, 28]. A range of possibilities has been postulated to explain their luminosity, such as the spin-down of a rapidly rotating young magnetar[29, 30, 31], the interaction of the SN ejecta with a massive (3–5 M☉) C/O-rich circumstellar medium[32, 33], an explosion via pair instability supernova[34, 35] or the consequences of a pulsational pair instability event[36, 37, 38, 39].

**1.1 - How do we define a superluminous supernova?**

Due to the incredible discovery pace of SLSNe, especially the hydrogen-free type which have been observed ten times more than the hydrogen-rich, we have roughly 110 candidates at the time of writing, spanning from those in neighbouring galaxies (SN2018bsz at 111 Mpc)[40] up to the distant Universe (DES16C2nm at $z$~2.0)[41]. Thanks to such a wealth of discoveries, augmented by the advent of all-sky surveys[42, 43, 44], the original concept of using an arbitrary fiducial luminosity threshold to define such a population[17] has now been reconsidered. From large sample studies it seems that the SLSN I luminosity function extends down to the luminosities of broad-line type Ic SNe ($M_{AB}$ ~ −20 mag)[42] or even fainter at the level of a more prosaic type Ic SN ($M_{AB}$ ~ −19 mag)[44]. Nevertheless, such samples are very heterogeneous in how they were selected, focusing on slow-rising light-curves or the presence of O II lines around 4000−4400Å in their pre-peak spectra. The latter, although a powerful signature, it is not unique of SLSNe (e.g. type Ib SN2008D)[45] and it is a consequence of non-thermal excitation of CNO layers[46] at a temperature between 12,000K and 15,000K[25]. Hence, the luminosity function tail towards dimmer magnitudes can be a consequence of the selection criteria. Ideally, to have a phenomenological definition similar to those of other SN types, we would need the spectral evolution from before peak luminosity up to 30 days after. Indeed, a SLSN I spectrum at 30 days resembles that of a type Ic at peak[22, 27], exhibiting a photospheric velocity that does not evolve after +30 days in contrast with typical stripped-envelope SNe[47, 48, 49, 50].

SLSN I are spectroscopically distinct from other types of SNe (i.e. SNe II/Ia/Ib), although a single feature that is always present in all SLSNe I but never in normal-luminosity SNe Ic or vice versa, has not been found[49]. However, at present, observations of the majority of objects do not cover such spectroscopic evolution and the advent of the Large Synoptic Survey Telescope (LSST) or satellite surveys, such as Euclid, with their hundreds of SLSN discoveries (ref. [51] and Dark Energy Survey collaboration, personal communication) will make it more challenging. To solve such issue a method has been proposed[48], using machine learning algorithms, to statistically describe and identify SLSNe I from their multi-band photometric behaviour in a four-dimensional parameter space (i.e. using four photometric parameters to select SLSN I). This method requires the light-curve of the evolution in two bands up to 30 days after maximum light and only a single spectrum in the same timeframe to confirm the resemblance to well-studied SLSNe I.

**1.2 –Spectrophotometric evolution of superluminos supernovae**

Originally, a possible division has been proposed[17] among SLSNe to create two subclasses of hydrogen-free event: rapidly declining (SLSNe I) and slowly declining (SLSNe R, named after a resemblance of their decline with that of the radioactive decay of $^{56}$Co), together with the hydrogen-rich events with the luminosity evolution driven by an interaction with a massive CSM. Although the hydrogen-free subdivision was not well received for several years[47], analyses based on detailed studies of small datasets[52], large spectroscopic datasets[49] and statistical approaches[48] have now supported such bimodality. Considering that the hydrogen-rich events can be divided into two subclasses, dominated by interaction and not, four types of SLSNe exist: SLSN I Fast (or F) for which SN2011ke[22, 49] could be considered the prototype; SLSN I Slow (or S) for which SN2015bn[53,54] is the best observed example; SLSN II with SN2013hx[21] as the archetype and SLSN IIn dominated by interaction as exemplified by SN2006gy[18]. In the following, SLSNe IIn will not be addressed since their spectrophotometric properties are very similar to standard SNe IIn. SLSN light-curves have long timescales with an average rise of 28 and 52 days for SLSNe I Fast and Slow, respectively[47, 55] (see Table 1 for further information), whereas SLSNe II show an average of 34 days rise time[21], although that is based on a very small sample. The rise time of SLSN I Slow can reach up to ~100 days (e.g. PS1-14bj)[56] and they show a hot (14,000 to 22,000 K)[57, 58] early-time luminosity excess[55, 57, 58, 59] (see Fig. 2), referred to as 'bump'[59]. It has been speculated that such a bump could be ubiquitous for all types of SLSN I[60], but this has been recently disproved[44]. The overall evolution of the prototypical SLSN I Slow SN2015bn revealed the presence of undulations in the light curve, possibly due to the interaction of the expelled material with small clumps of CSM[53]. A close inspection of the post-peak light-curve behaviour of a sample of nearby SLSNe I Slow revealed the presence of such undulations in all of them[52], a conclusion also supported by the analysis of a larger sample of Palomar Transient Factory (PTF) SLSNe[42]. Although it is not clear if such undulations are ubiquitous for this subclass, they seem a characteristic trademark since they have not been observed in SLSNe I Fast[42, 52], where only two objects show a small rebrightening, SSS120810[61] and PS16aqv[62] (although for the second event the phenomenon is only observed in *g*- and *i*-band and not in *r*-band). As could be expected, SLSN I Slow light-curves have been observed up to later phases than the Fast[42, 43, 52, 54, 55, 63] and, as displayed in Fig. 2, show an initial increase in the decline at ~150 days and then a steeper one from 300 to 400 days following a power law of $t^{-5}$ (ref. 52), which is an identical slope to that of an adiabatic expansion phase (the Sedov–Taylor phase) and it has been observed in all the few SLSNe followed until such phase. After that, the light-curves slowly decline and become similar to what is experienced by SLSN I Fast after 100 days post-peak, although such timeframe has only been explored for the nearby SN2015bn[54].

Conversely, SLSN I Fast light-curves are smoother with no observed 'bump' but with a characteristic 'tail' displayed after roughly 50 days[22, 61, 64] which might be due to the most favoured powering mechanism[22], i.e. a rapidly rotating magnetar[29, 30, 31, 65]. Their light-curve around peak evolution shows a decline slower than the rise by a factor ~2, similar to what experienced by SLSN I Slow[42, 47]. The post-peak evolution appears to be faster for fainter events, different to what is experienced by SLSN I Slow, where the decline rate is roughly the same regardless of peak luminosity[7, 42, 48].

The shape of the SLSNe II light-curve around peak epoch is asymmetric and the decline tends to be slower than that of Fast SLSNe I due to hydrogen recombination[19, 20, 21], and faster than that of Slow SLSNe I after 40 days. Noticeably, all events with late-time data (>150 d) show a flattening of the light-curve due to interaction with a hydrogen-rich shell previously expelled[21], although the numbers are too low to draw any robust conclusion.

From a spectroscopic point of view, pre-peak SLSNe show blue, almost featureless, spectra with few lines due to highly ionised elements. During this pre-peak phase, spectra show blackbody temperatures from 22,000K to 12,000K (refs. [46, 58]). Ions such as C III, C II, Ti III, Mg II and Si II dominate the ultraviolet part of SLSN I spectra with a broad absorption from 2,200 to 2,600Å up to soon after peak epoch[41, 49, 66, 67, 68, 69, 70, 71]. In the optical region of the electromagnetic spectrum, SLSNe I Fast are mainly dominated by the typical O II lines[6, 49, 58] (Fig. 3), while SLSNe I Slow also show Si II and Fe III features, as well as Fe II and O I as soon as the ejecta temperature approaches 12,000K (ref. [53]).

The ejecta velocity is quite similar for all ions and objects[10, 49] with lines from single transitions showing narrow velocity widths (1,500 km s$^{-1}$)[10], making SLSNe I expansion velocity and velocity dispersion (line width) quite different[10] than what observed in SNe Ib/c[72]. At 10 days after peak, SLSNe I (Fast and Slow) have average Fe II λ5169 absorption velocities of -15,000 ± 2,600 km s$^{-1}$ at 10 days after peak, which are higher than those of SNe Ic by ∼7000 km s$^{-1}$ on average[50].

SLSN II are instead featureless around peak epoch with a broad, very shallow Hα profile[19, 20, 21]. Soon after that, at around 10 days after peak appears the He II λ4686 line, which is observed for roughly a week[19, 20]. Recently, observations of some SLSNe I Slow have shown interaction with a hydrogen-rich shell at ≳100 days from maximum light[73, 74]. Noticeably, all of these events show strong C II lines in pre-peak to peak spectra instead of, or with weaker, O II lines[40, 74]. In addition, light-curves of these objects may differ from the prototypical evolution due to the presence of multiple peaks (iPTF15esb)[74] or a 'plateau' phase during the rise to the peak (SN2018bsz)[40].

A second key window in the spectral evolution is when SLSNe have spectra similar to their normal luminosity counterparts, but somewhat delayed (30 days for SLSNe I[27] and 20 for SLSNe II[21]), and temperatures have cooled down to 7,000K < $T$ < 10,000K. At this epoch the strongest features are Ca II H&K lines, several Fe II multiplets around 5000 Å and Mg I] λ4571, which tend to appear after 35 days[22, 53]. In SLSNe I Slow the region between 5,000 and 5,800 Å shows a broad emission feature that is likely a blend of Mg I λ5180, [Fe II] λ5250 and [O I] λ5577, while for SLSNe I Fast such emission is shallower and possibly due to a forest of Fe II lines[22, 27, 50, 52, 53]. Forbidden lines such [O I] λλ6300, 6363 and [Ca II] λλ7291, 7323 become visible at this phase in SLSNe I Slow, whereas they never appear in SLSNe I Fast with the exception of the event Gaia16apd[64], which however is an intermediate event between the two subclasses, and the only one so far. Soon after 30 days, the spectral evolution of SLSNe I Fast and Slow freezes and there are no noticeable changes in the observed elements, with only some line profiles changing in strength[48, 52, 53].

SLSNe II show the typical elements observed in bright, linear declining type II (or type IIL, see Modjaz, Arcavi and Gutierrez review[75]) such as Fe II multiplet λλ4924, 5018, 5169, Na ID and Balmer lines, as well as a high-velocity Hα feature[76, 77]. Generally speaking, during the photospheric phase, SLSNe II temperatures derived from blackbody fits are slightly lower and have a slower evolution with respect to those of SLSNe I[21].

At late times (>150d) a SLSN enters into what is defined a nebular phase, dominated by forbidden emissions representing the spectral fingerprints of the supernova's deep interior, giving a unique opportunity to see what an exploded star looks like inside. In truth, no nebular spectrum of SLSNe I Fast has been observed, with the exception of the transitional case of Gaia16apd[64, 78] and, as of June 2019, the latest available spectrum with signal-to-noise > 5 has been observed at 121 days after peak[49] (Fig. 3). Such a spectrum does not show any evolution from earlier spectra at 30–50 days and no forbidden transitions in emission. SLSNe I Slow also show the same features they developed at 30–50 days but forbidden lines such as [O I] and [Ca II] are now stronger. The [Ca II] line arises earlier in SLSNe I Slow compared to SLSNe I Fast, suggesting that the differences between the two subclasses are not mainly due to different ejecta masses as once proposed[47]. The centroids of such lines have been shown to move in time. Ionized elements show line profiles distinct from the neutral ones and that are likely from a region interior to that where the neutral lines are formed, and where occultation effects can be easily produced[52]. This suggests that multiple emitting regions are responsible for the overall spectral features of SLSNe I Slow[52, 78, 79]. A strong Ca II triplet, O I λ9263, O I 1.13 µm and Mg I 1.50 µm are also visible but no distinct He, Si, or S emission is[79]. SLSNe I that showed a strong C II signature around peak epoch systematically display a Hα profile, sometimes with a few components (ePESSTO collaboration, personal communication), due to the interaction with a hydrogen-rich CSM[71, 73]. A similar profile is also displayed in SLSNe II where the ejecta collides with a ring-shaped, or clumpy, CSM[21]. Overall, only SLSNe I Slow show nebular spectra and they display little to no variation in the observed elements from 40 days up to roughly 1,000 days[52,54]. The nebular spectra of this subclass show strong similarity with broad-line type Ic SNe such as SN1998bw and can be reproduced by models with an oxygen-zone of $M \gtrsim 10 M_\odot$, pointing to an explosion of a massive CO core requiring a zero-age main sequence mass of $M_{ZAMS} \gtrsim 40 M_\odot$ (ref. [79]).

**1.3 – X-ray, radio and polarisation signatures of superluminous supernovae**

SLSNe I have been the target of an extensive campaign to locate a high-energy counterpart[52, 80, 81, 82] spanning from some days after the alleged explosion epoch up to roughly 2,000 days after[81]. Only two SLSNe I, out of roughly 30, show an X-ray signature: the Fast SCP06F6 ($L_x \sim 10^{45}$ erg s$^{-1}$)[80] and the Slow PTF12dam ($L_x \sim 10^{40}$ erg s$^{-1}$)[81]. The fact that the majority of the X-ray limits lie at $10^{40} < L_x$ (erg s$^{-1}$) $< 10^{45}$ [52, 80, 81] suggests that the X-ray emission associated with SCP06F6 is not common among SLSNe I[81]. The SLSNe I X-ray limits would point toward a magnetar magnetic field and ejecta masses similar to what is derived from light-curve models (3 M$_\odot \lesssim M \lesssim$ 30 M$_\odot$)[22, 29, 43, 53, 55, 61, 83], and would disfavour CSM interaction as the main source of energy powering the light-curve[52, 81].

Searches at radio wavelengths for sources coincident with the location of SLSNe I have been conducted for a handful of cases but the majority of them has only provided limits $10^{27} <$ L$_\nu$ (8 GHz) $< 10^{30}$, where L$_\nu$ (8 GHz) is the luminosity at 8 GHz (refs. [53, 84, 85, 86]). Only the Fast PTF10hgi, which is the only one showing P-Cygni profiles of Hα and He I at around 50 days from peak brightness[49], has been observed at radio frequency about 7.5 years post-explosion, with a luminosity of L$_\nu$ (6 GHz) $\approx 1.1 \times 10^{28}$ erg s$^{-1}$ Hz$^{-1}$ (ref. [86]). Radio limits disfavour a radio emission similar to that of faint un-collimated GRBs and relativistic supernovae[85], while the only observation is reminiscent of the quiescent radio source associated with a repeating FRB and favour a central engine powered nebula as a source of the observed luminosity.

Spectropolarimetric observations of SLSNe I have been reported only for two SLSNe I Slow, SN2015bn[87] and SN2017egm[84], although data have been collected for other two Slow events (G. Leloudas and M. Bulla, personal communications). Both observations showed an increase in the polarisation from pre-peak to post-peak epoch, also supported by imaging polarimetric analysis[88]. For SN2015bn, the polarization spectrum is characterised by a dominant axis and by a strong wavelength dependence, likely due to line opacity. A Monte Carlo model suggests that such properties can be reproduced by an axisymmetric ellipsoidal configuration for the ejecta. The increase of polarization can be a consequence of the increase in the asphericity of the inner layers of the ejecta or the fact that the photosphere recedes into less spherical layers[87]. Imaging polarimetric data have also been reported for the SLSN I Fast LSQ14mo[89] and for the SLSN II PS15br[21], but the former does not present significant asymmetries, while the latter shows some degree of polarisation that, however, cannot be deemed conclusive due to unaccounted-for sources of scattering[21].

**1.4 – Closing remarks on nature's brightest fireworks**

It is beyond the scope of this review to focus on the energy source of such objects but the magnetar scenario seems able to reproduce all the main observables, although some hybrid models (i.e.

magnetar + interaction) might be equally or more effective[10,65]. Nevertheless, a small to medium degree of interaction is observed in all SLSNe I Slow from light-curve behaviour[52, 53, 55, 58, 59, 84] or spectroscopic features[52, 56, 70, 71, 90]. It is also noteworthy that SLSNe I Fast can be used to constrain cosmological parameters up to $z\sim2$ (Dark Energy Survey collaboration personal communication).

## 2 – Fast Blue Optical Transients

The field of rapidly evolving transients has been flourishing during the last years thanks to the advent of wide wide-field surveys like PanSTARRS, the Dark Energy Survey (DES), the Asteroid Terrestrial-impact Last Alert System (ATLAS) and the Zwicky Transient Facility (ZTF), which produced a decisive burst in the active investigation of such transients[8,9,91]. However, the plethora of objects that can fit such broad observational definition makes it difficult to have a clear focus on their nature even with the approach highlighted in Fig. 1. In such a description could be fitted objects linked to the large explosion of a massive star with very small ejected mass, such as SNe 2010X, 2002bj and 2005ek[92, 93, 94]. Transients likely related to a thermonuclear explosion and known as 'Ca-rich objects'[95] and those consistent with a helium shell detonation on a white dwarf (known as '.Ia')[96], such as OGLE-2013-SN-079[97], could also be covered by the above definition. Some interacting type Ibn supernovae might also fit the rapidly evolving transients paradigm[98, 99]. Nevertheless, despite these objects showing a fast rise and decline (see Fig. 4 for some examples), the majority of them can (and have been) explained with the final outcome of a massive stripped star or a thermonuclear explosion (Review Articles by Jha, Maguire and Sullivan on thermonuclear SNe[100] and Modjaz, Gutierrez and Arcavi on core-collapse SNe[75]) as also hinted by their position in the phase-luminosity diagram (Fig. 1), which is below the lines representing the maximum possible luminosity from a standard SN explosion. Some of these transients overall evolution is usually not as fast as the events presented in the Pan-STARRS (or PS1), the Subaru Hyper Suprime-Cam Transient Survey and the Dark Energy Survey (DES) samples[8, 9, 91], which in literature are becoming more often referred to as Fast Blue Optical Transients (FBOTs). Moreover, the transients of such samples, as well as the nearby event AT2018cow at 60 Mpc[101, 102, 103, 104, 105], show timescales and spectroscopic evolution inconsistent with any standard scenario or model able to reproduce the observables (or some of them) of the objects mentioned above.

**2.1 – Characteristic features of FBOTs**

FBOTs are usually characterised by a rapid light-curve rise to the peak (≲10 days from the last non-detection) and an exponential decline in 30 days after peak[9] (Fig. 4, left panel), or a time above half-maximum luminosity ($t_{1/2}$) of less than 12 days[8] (Fig. 4, right panel). They are usually quite blue at peak (g-r < -0.2) and slowly become redder. Their spectral energy distribution (SED) at peak can be fitted with hot blackbodies (10,000 – 30,000K) and an optically thick ejecta[8, 9, 106] showing photospheric temperatures/radii that cool/expand with time, which is what expected in case of a supernova explosion. Although KSN2015K has mainly a single band light curve[107], the single-epoch colour suggests an equally blue transient. A similar behaviour is shown for SLNS04D4ec[106] and by the nearby AT2018cow, which display a 30,000K temperature at peak epoch, slowing decreasing to roughly 17,000K after 30 days[101, 102, 103]. The ROTSE-IIIb transient named 'Dougie' could also fit in such description since it is blue, almost featureless and fast evolving[108] (see Fig. 4). Nevertheless, its spectral sequence and overall observables seem to favour more a TDE interpretation than the other FBOTs. While this review was under evaluation, another fast-riser object (SN2018gep) and a fast-decliner transient (SN2019bkc) have been presented in submitted papers[109, 110]. The first object shows a rise of 4 days in r-band, a decline of 13 days to reach the above half-maximum luminosity, which is borderline with the above definition, and it is spectroscopically classified as a broad-lined stripped-envelope supernova[109]. The latter event shows a steep decline in the first 10 days after peak luminosity[110]. Moreover, at least two other objects fitting such photometric behaviour have been recently discovered (O. McBrien, D. Perley, S.

Prentice, S. J. Smartt personal communication) by ATLAS and ZTF, suggesting that more will come in the future.

The peak magnitudes of FBOTs span from the fainter end of core-collapse SNe up to luminosities comparable to those of SLSNe without any significant trend between peak luminosity and evolution[7]. Their rise time is faster than the decline and the redder the band, the longer is the exponential decline timescale[8, 9]. The post-peak decline timescales observed in the PS1 and DES samples cover a wide range, which is difficult to explain with a single value for an exponential decline, or in general with a single scenario or powering mechanism[9]. It is interesting to note that the declines of AT2018cow and KSN2015K become slower after ~15 days[102, 107], similar to what happens to some of the PS1 objects[8]. This might hint of a change in the spectrophotometric properties[102] or in the powering mechanism responsible for the light curve[105].

From a spectroscopic point of view, these transients are hot, featureless blackbodies at peak (no pre-peak spectroscopic observation has been carried out at the time of this Review, see Fig. 5), with a lack of narrow permitted lines in emission typical of interacting SNe[8]. Nevertheless, they resemble more events powered by interaction or recombination rather than by radioactive decay[8, 9]. Of the PS1 and DES samples, only PS1-12bv shows a broad spectral feature near 3,900 Å and tentative evidence for CSM absorption linked to the Mg II $\lambda\lambda$ 2796, 2803 feature[8].

Conversely, around its peak AT2018cow exhibits a broad-like feature at ~4,600 Å[102]. This feature is vaguely reminiscent of a Fe II line usually observed in type Ic SNe at this phase, although in this case the broad absorption disappeared within few days. Weak, redshifted emission of hydrogen and helium, both ionised and neutral, appears after ~10 days[101, 102, 103] (see Fig. 5, right panel). These become more prominent, centred and with a red shoulder after 35 days[102, 103] up to at least 85 days[103]. As recently pointed out[105], such features resemble those of type Ibn, although in that SN class these features are already seen around peak. Moreover, type Ibn events show a Fe II pseudo-continuum at wavelengths bluer than 5,400 Å[111], which is not observed in other FBOTs. Among the heterogeneous variety of type Ibn[98, 99], AT2018cow late-time spectra show the greatest resemblance to SN2011hw, a transitional type IIn/Ibn SN[112, 113], which is also the only Ibn with a time above half-maximum luminosity typical of the FBOTs.

X-ray monitoring of AT2018cow[103, 114], the only transient among the FBOTs with X-ray information, revealed a hard X-ray spectral component at $E \geq 10$ keV and luminous and highly variable soft X-rays, with properties unprecedented among astronomical transients[103]. An abrupt change in the X-ray decay rate and variability appears to accompany the change in optical spectral properties, while the bright radio emission detected is consistent with the interaction of a blast wave with almost relativistic velocities in a dense environment[103].

Due to the unusual observational properties observed in such transients, a wide range of possibilities has been suggested to explain their behaviour. The PS1 and DES samples showed blue, continuum dominated, spectra consistent with what observed for transients powered by a shock-breakout or recombination in an extended envelope[8] that might be due to an optically thick, low mass circumstellar wind surrounding a core-collapse SN[9]. Such an explanation is also the most widely used since can account for the variety of observed luminosities and timescales displayed by the FBOTs. However, the wealth of data gathered for AT2018cow opened up more possibilities. For example, several suggested scenarios to explain the object are linked to a central X-ray engine model (ref. [103] and reference therein), e.g. a failed explosion of a blue supergiant with the formation of a black hole or an electron-capture SN with a millisecond magnetar. Other explanations point towards a tidal disruption event from an intermediate mass black hole (refs. [102, 104] and references therein); formation of a magnetar in a binary neutron star merger (ref. [101] and references therein) or interaction with a CSM[103, 105, 114]. However, some of the above power sources/progenitor scenarios rely on the notion of an X-ray source which has only been observed for AT2018cow and hence might not be representative of or valid for the whole class.

# 3. The future of extreme supernovae

The second decade of observations of such extreme supernovae will likely be more thriving than the first and the achievements will hopefully surpass those already accomplished. Thanks to the advent of next-generation telescopes - such as LSST, Euclid and the James Webb Space Telescope - SLSNe will be observed in a far greater number with a longer redshift baseline up to distances previously inaccessible, opening a new era of high-redshift probes and cosmology. High-resolution spectroscopic facilities, such as the Very Large Telescope (VLT) with X-SHOOTER or UVES (Ultraviolet and Visual Echelle Spectrograph) and the Keck telescope with HRIS (High Resolution Echelle Spectrometer), fed by high-cadence surveys such as ATLAS and ZTF will provide many more FBOTs with late-time data. This will unravel any connection between such an observed class and the final stages of massive stars.

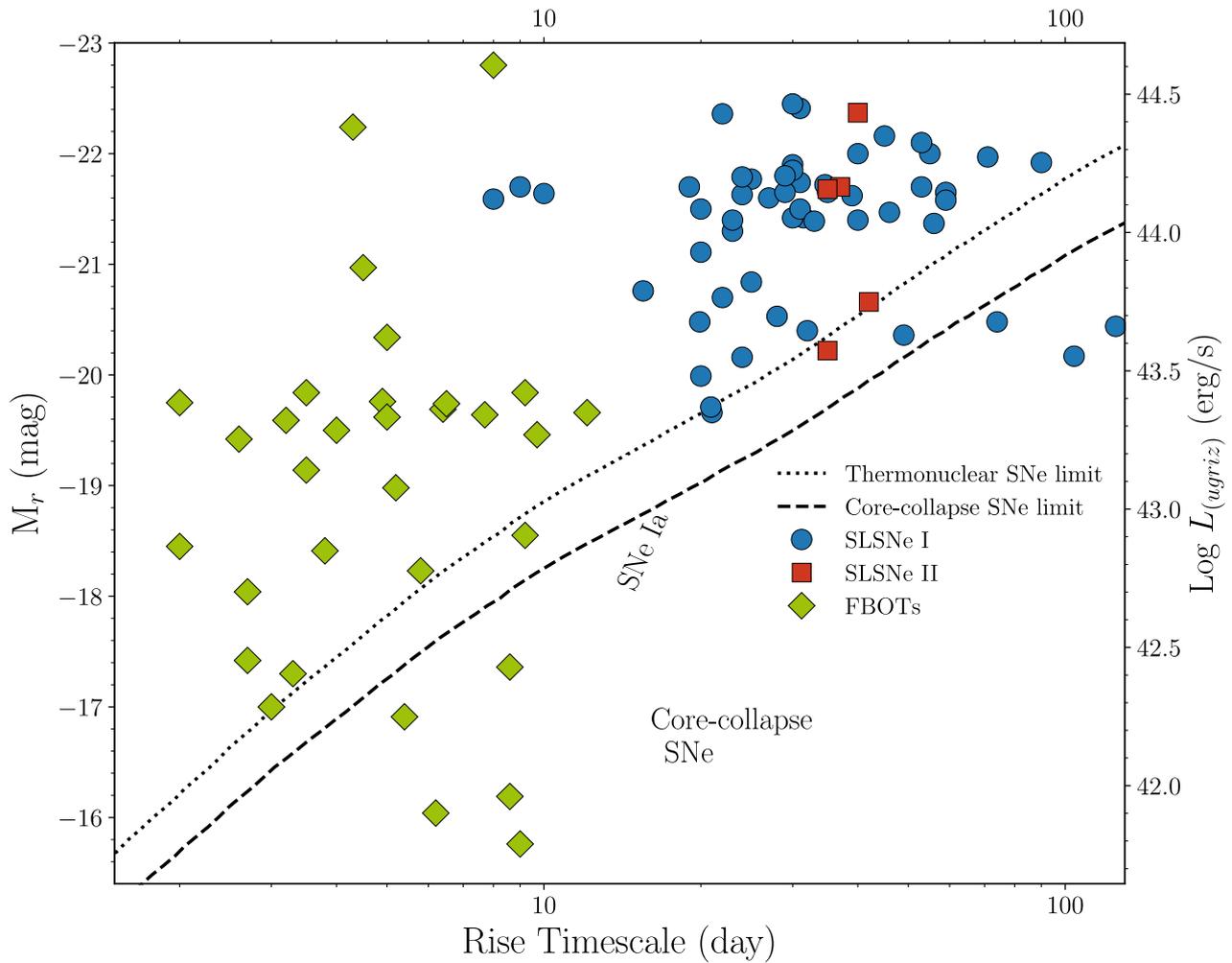

**Fig 1 / The transient parameter space with peak luminosity as a function of the rise time.** The left y-axis is the r-band peak luminosity in units of absolute magnitudes and the right y-axis is the pseudo-bolometric luminosity. The extreme transients that are the focus of this Review Article lie above the lines representing the maximum luminosity, as a function of the rise time, that a core-collapse (dashed line) or thermonuclear (dotted line) event can produce. The redshift distribution for SLSNe is $0.03 \lesssim z \lesssim 2.0$, while fast transients (FBOTs) have been observed from $z = 0.014$ up to $z = 1.56$, spanning over 6 mag.

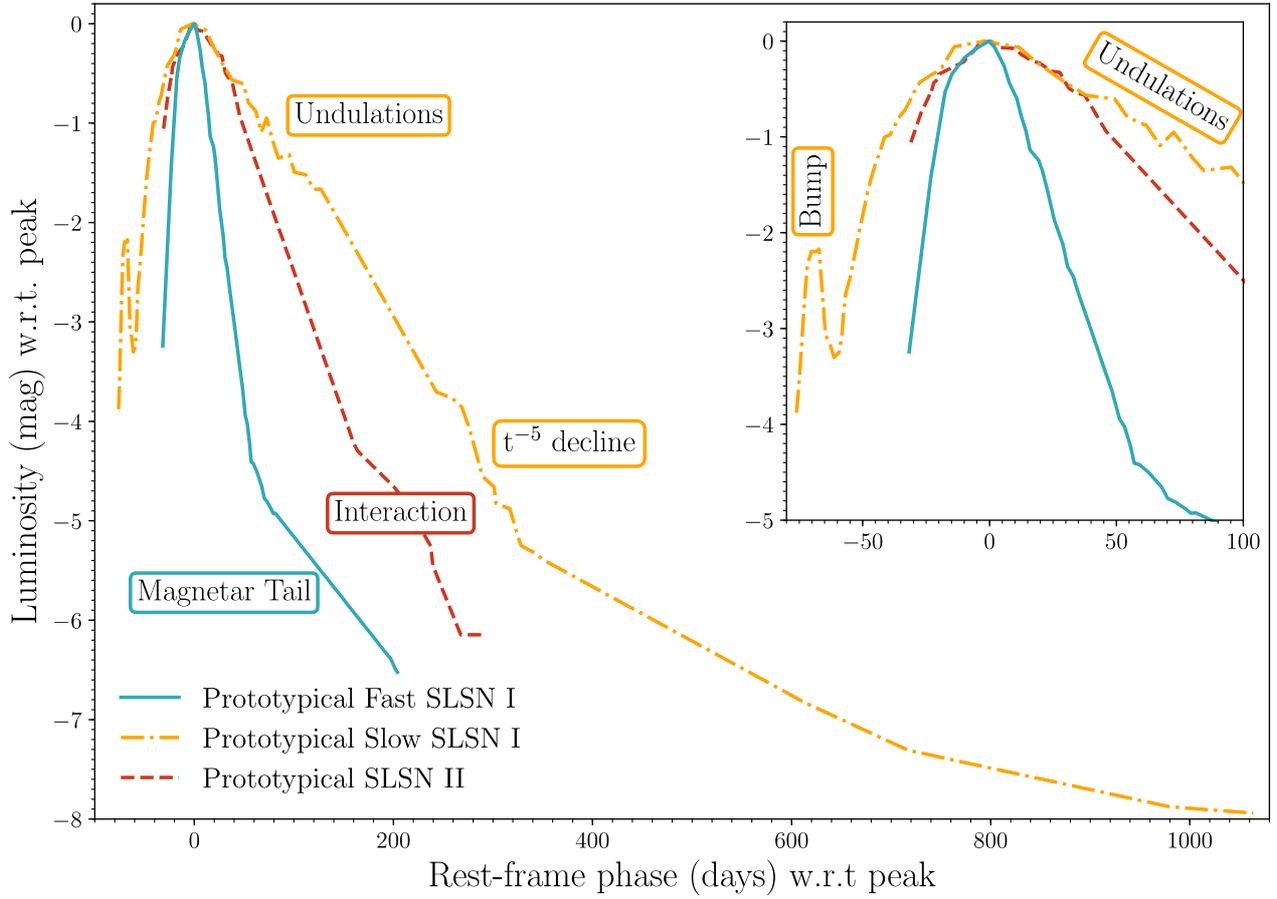

**Fig. 2 | Prototypical light-curve evolution of SLSNe I and II.** The luminosity with respect to (w.r.t.) the peak is plotted against the rest-frame phase w.r.t. the peak. The main observables of the evolution are noted, together with a zoom of the peak phase in the top right. The width and length of each phase (i.e. bump, peak, undulation, $t^{-5}$ decline and tail) might differ from object to object, while the Slow SLSN I behaviour after 400 is mainly driven by SN2015bn[54]. In the insert, the bump and undulations observed in SLSN I Slow are noticeable. These prototypical light-curves constructed with data from refs. [**19, 20, 21, 22, 27, 52, 53, 54, 58, 59, 61, 83**].

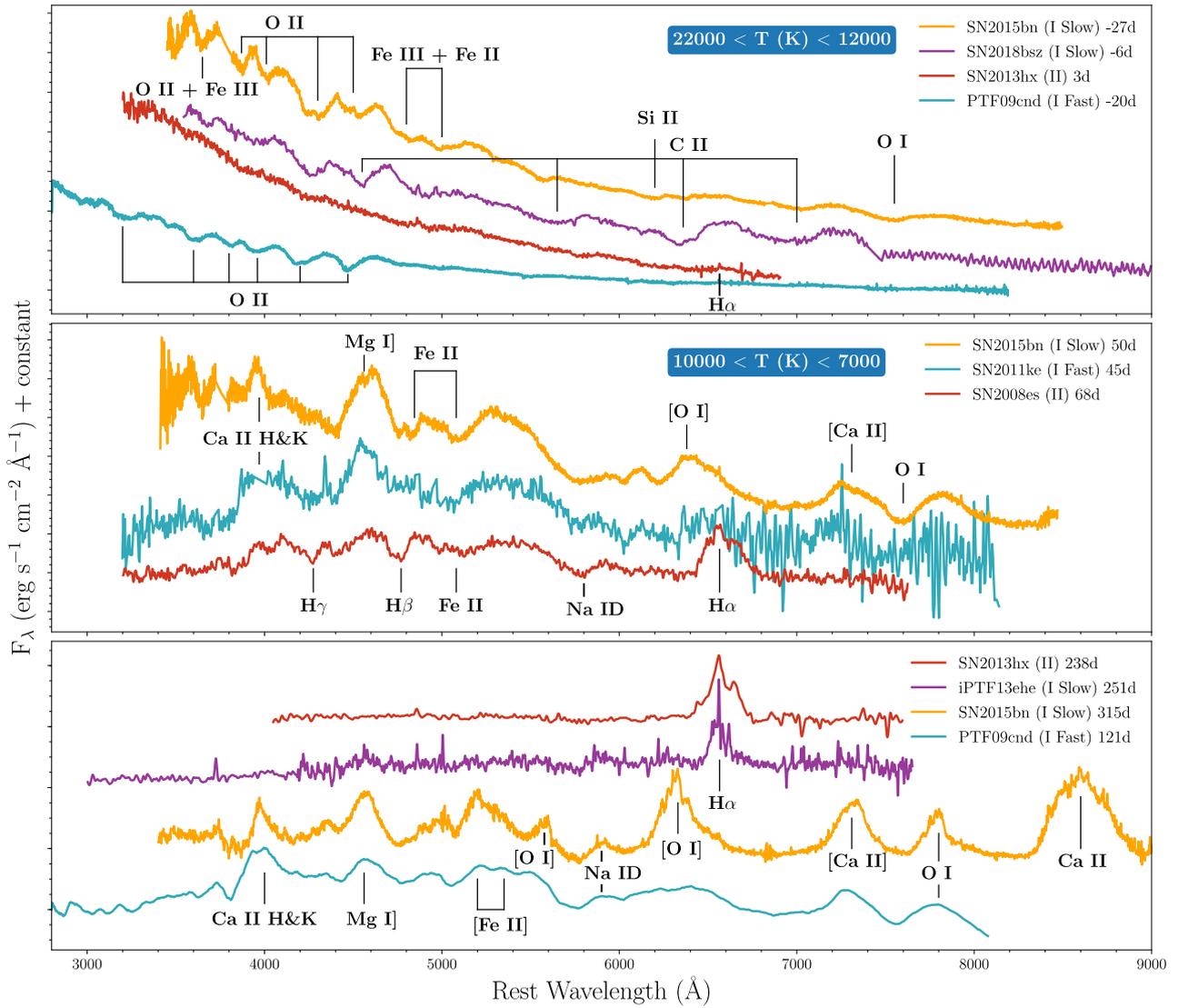

**Fig. 3 | Spectroscopic evolution of the three SLSN classes.** The evolution can be divided into three key time windows: up to the luminosity peak (top panel); around 50 days, when the ejecta velocity is frozen (middle panel), late-time (>100 d) where nebular emissions might be seen (bottom panel). Key spectral features and temperature range are reported. Data from refs. [**6, 19, 21, 22, 40, 49, 53, 74, 79**].

|  | **SLSNe I Fast** | **SLSNe I Slow** | **SLSNe II** |
|---|---|---|---|
| Early bump | No | Yes | No |
| Light-curve typical rise time; lower-upper limits (day) | 28 (13-35) | 52 (33-100) | 34 (31-36) |
| Light-curve undulations | No | Yes | No |
| Spectra coverage (day) | -20 < phase < 121 | -28 < phase < 1080 | 3 < phase < 340 |
| X-ray detection | Yes | Yes | No |
| Polarimetry information | Imaging | Imaging and spectropolarimetry | Imaging |
| Polarization | No | Yes | Yes but inconclusive |
| Nebular spectroscopy | No* | Yes | Yes |
| Late-time H$\alpha$ emission | Yes | Yes | Yes |

**Table 1 | Observables of superluminous supernova subclasses.** *The only SLSN I Fast with nebular spectroscopy is Gaia16apd, which is a transitional case between the Fast and the Slow.

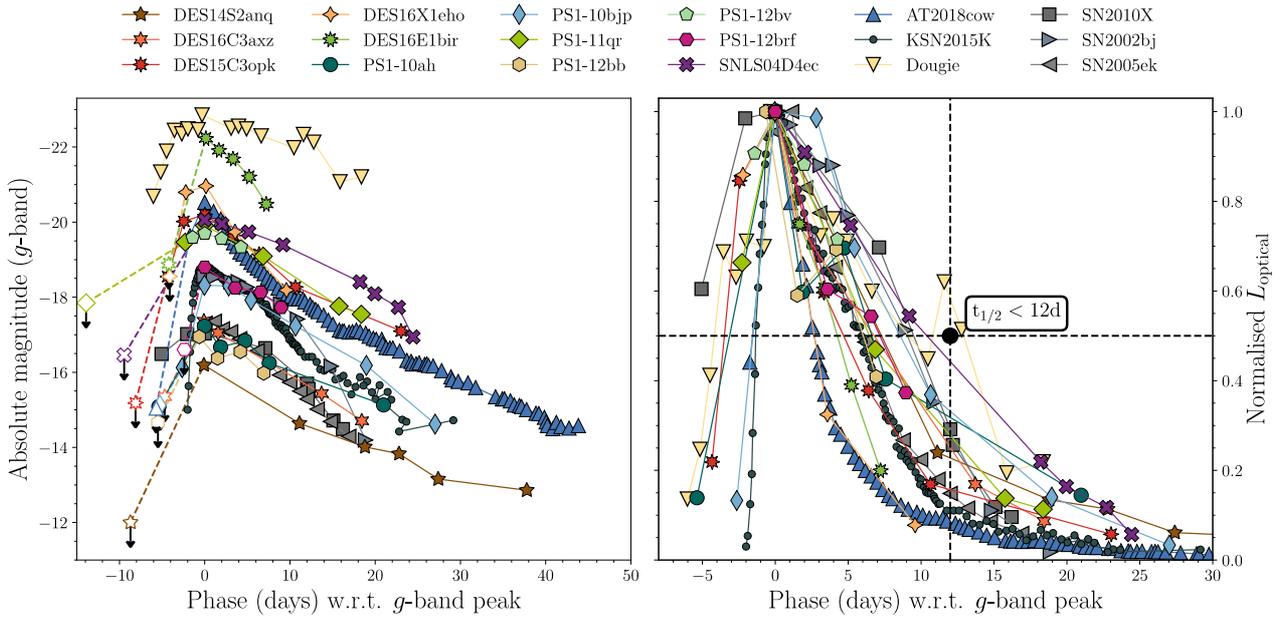

**Fig. 4 | Light-curve evolution of known rapidly evolving transients (or FBOTs).** Data are taken from the Pan-STARRS survey (PS1) [8], Dark Energy Survey (DES) [9], Supernova Legacy Survey (SNLS) [106] and the Kepler supernova program [107]. Additional data on the best-observed object AT2018cow, the luminous event 'Dougie' and other fast transients linked to the 'Ca-rich' family (in grey) are added [92, 93, 101, 102, 108, 115].

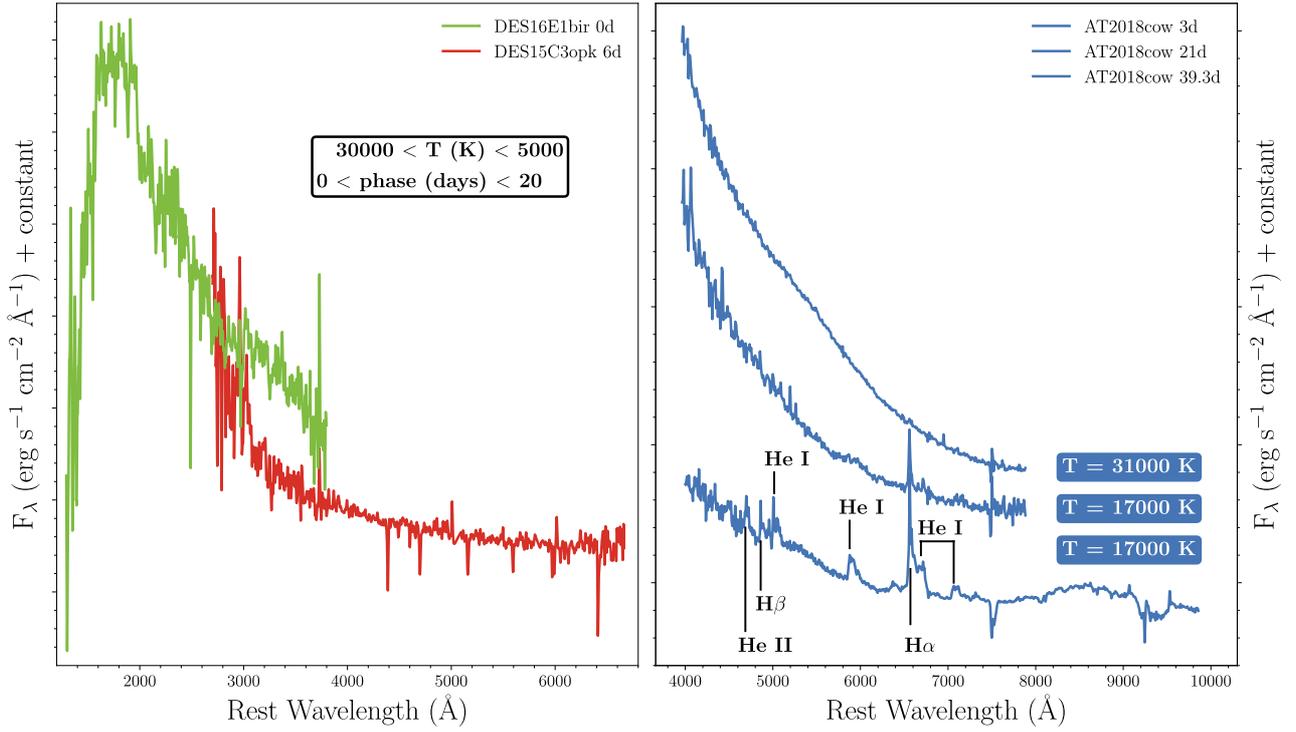

**Fig. 5 | Blue, featureless spectra of Fast Transients (or FBOTs).** Left: two examples of FBOT spectra together with the temperature range measured by the DES and PS1 FBOTs sample over the first 20 days after peak luminosity. Right: a snapshot of the spectroscopic and temperature evolution of the best-observed transient, AT2018cow. Data from refs. [9, 101, 102].

**References:**

1. Janka, T. Explosion Mechanisms of Core-Collapse Supernovae. *ARNPS* **62,** 407-451 (2012).
2. Hillebrandt, W. & Niemeyer, J. C. Type Ia Supernova Explosion Models. *Annual Review Astron. Astrophys.* **38,** 191-230 (2000).
3. Arnett, W. D. Type I supernovae. I - Analytic solutions for the early part of the light curve. *Astrophys. J.* **253,** 785-797 (1982).
4. Umeda, H. & Nomoto, K. Nucleosynthesis of Zinc and Iron Peak Elements in Population III Type II Supernovae: Comparison with Abundances of Very Metal Poor Halo Stars. *Astrophys. J.* **565,** 385-404 (2002).
5. Childress M. et al. Measuring nickel masses in Type Ia supernovae using cobalt emission in nebular phase spectra. Mon. Not. R. Astron. Soc. **454,** 3816-3842 (2015).
6. Quimby R. et al. Hydrogen-poor superluminous stellar explosions. *Nature*, **474,** 487-489 (2011).
7. Inserra, C. & Smartt, S. J. Superluminous Supernovae as Standardizable Candles and High-redshift Distance Probes. *Astrophys. J.* **796,** 87 (2014).
8. Drout, M. et al. Rapidly Evolving and Luminous Transients from Pan-STARRS1. *Astrophys. J.* **794,** 23 (2014).
9. Pursiainen, M. et al. Rapidly evolving transients in the Dark Energy Survey. *Mon. Not. R. Astron. Soc.* **481,** 894-917 (2018).
10. Gal-Yam, A. The Most Luminous Supernovae. Preprint at arXiv:1812.01428 (2018).
11. Lunnan, R. et al. Hydrogen-poor Superluminous Supernovae and Long-duration Gamma-Ray Bursts Have Similar Host Galaxies. *Astrophys. J.* **787,** 138-(2014).
12. Leloudas, G. et al. Spectroscopy of superluminous supernova host galaxies. A preference of hydrogen-poor events for extreme emission line galaxies. *Mon. Not. R. Astron. Soc.* **449,** 917-932 (2015).
13. Perley, D. et al. Host-galaxy Properties of 32 Low-redshift Superluminous Supernovae from the Palomar Transient Factory. *Astrophys. J.* **830,** 13-(2016).
14. Chen, T.-W. et al. Superluminous supernova progenitors have a half-solar metallicity threshold. *Mon. Not. R. Astron. Soc.* **470,** 3566-3573 (2017).
15. Schulze, S. et al. Cosmic evolution and metal aversion in superluminous supernova host galaxies. *Mon. Not. R. Astron. Soc.* **473,** 1258-1285 (2018).
16. Hatsukade, B. et al. Obscured Star Formation in the Host Galaxies of Superluminous Supernovae. *Astrophys. J.* **857,** 72-(2018).
17. Gal-Yam, A. Luminous Supernovae. *Science* **337,** 927 (2012).
18. Smith, N. et al. SN 2006gy: Discovery of the Most Luminous Supernova Ever Recorded, Powered by the Death of an Extremely Massive Star like η Carinae. *Astrophys. J.* **666,** 1116-1128 (2007).
19. Gezari, S. et al. Discovery of the Ultra-Bright Type II-L Supernova 2008es. *Astrophys. J.* **690,** 1313-1321 (2009).
20. Miller, A. et al. The Exceptionally Luminous Type II-Linear Supernova 2008es. *Astrophys. J.* **690,** 1303-1312 (2009).
21. Inserra, C. et al. On the nature of hydrogen-rich superluminous supernovae. *Mon. Not. R. Astron. Soc.* **475,** 1046-1072 (2018).
22. Inserra, C. et al. Super-luminous Type Ic Supernovae: Catching a Magnetar by the Tail. *Astrophys. J.* **770**, 128 (2013).
23. Prajs, S. et al. The volumetric rate of superluminous supernovae at z~1. *Mon. Not. R. Astron. Soc.* **464,** 3568-3579 (2017).
24. Quimby, R. et al. Rates of superluminous supernovae at z~0.2. *Mon. Not. R. Astron. Soc.* **431,** 912-922 (2013).
25. McCrum, M. et al. Selecting superluminous supernovae in faint galaxies from the first year of the Pan-STARRS1 Medium Deep Survey. *Mon. Not. R. Astron. Soc.* **448,** 1206-1231 (2015).

26. Cooke, J. et al. Superluminous supernovae at redshifts of 2.05 and 3.90. *Nature* **491,** 228-231 (2012).
27. Pastorello, A. et al. Ultra-bright Optical Transients are Linked with Type Ic Supernovae. *Astrophys. J*. **724,** L16-L21 (2010).
28. Angus, C. R. et al. A Hubble Space Telescope survey of the host galaxies of Superluminous Supernovae. *Mon. Not. R. Astron. Soc.* **458,** 84-104 (2016).
29. Kasen, D. & Bildsten, L. Supernova Light Curves Powered by Young Magnetars. *Astrophys. J.* **717,** 245-249 (2010).
30. Woosley, S. Bright Supernovae from Magnetar Birth. *Astrophys. J*. **719,** L204-L207 (2010).
31. Dessart, L. et al. Superluminous supernovae: $^{56}$Ni power versus magnetar radiation. *Mon. Not. R. Astron. Soc.* **426,** L76-L80 (2012).
32. Chevalier, R. A. & Irwin, C. M. Shock Breakout in Dense Mass Loss: Luminous Supernovae. *Astrophys. J.* **729,** L6 (2011).
33. Chatzopoulos, E. et al. Analytical Light Curve Models of Superluminous Supernovae: $\chi^2$ minimization of Parameter Fits. *Astrophys. J.* **773,** 76-(2013).
34. Gal-Yam, A. et al. Supernova 2007bi as a pair-instability explosion. *Nature* **462,** 624-627 (2009).
35. Kozyreva, A. et al. Fast evolving pair-instability supernova models: evolution, explosion, light curves. *Mon. Not. R. Astron. Soc*. **464,** 2854-2865 (2017).
36. Woosley, S. E., Blinnikov, S. & Heger, A. Pulsational pair instability as an explanation for the most luminous supernovae. *Nature* **450,** 390–392 (2007).
37. Sorokina, E., Blinnikov, S., Nomoto, K., Quimby, R. & Tolstov, A. Type I superluminous supernovae as explosions inside non-hydrogen circumstellar envelopes. *Astrophys. J.* **829,** 17 (2016).
38. Tolstov, A. et al. Pulsational pair-instability model for superluminous supernova PTF12dam: interaction and radioactive decay. *Astrophys. J.* **835,** 266 (2017).
39. Woosley, S. E. Pulsational pair-instability supernovae. *Astrophys. J.* **836,** 244 (2017).
40. Anderson, J. P. et al. A nearby super-luminous supernova with a long pre-maximum "plateau" and strong C II features. *Astron. Astrophys*. **620,** A67-(2018).
41. Smith, M. et al. Studying the Ultraviolet Spectrum of the First Spectroscopically Confirmed Supernova at Redshift Two. *Astrophys. J.* **854,** 37(2018).
42. De Cia, A. et al. Light Curves of Hydrogen-poor Superluminous Supernovae from the Palomar Transient Factory. *Astrophys. J.* **860,** 100 (2018).
43. Lunnan, R. et al. Hydrogen-poor Superluminous Supernovae from the Pan-STARRS1 Medium Deep Survey. *Astrophys. J.* **852,** 81 (2018).
44. Angus, C. R. et al. Superluminous Supernovae from the Dark Energy Survey. Preprint at arXiv:1812.04071-(2018).
45. Soderberg, A. et al. An extremely luminous X-ray outburst at the birth of a supernova. *Nature* **453,** 469-474 (2008).
46. Mazzali, P. et al. Spectrum formation in superluminous supernovae (Type I). *Mon. Not. R. Astron. Soc.* **458,** 3455-3465 (2016).
47. Nicholl, M. et al. On the diversity of superluminous supernovae: ejected mass as the dominant factor. *Mon. Not. R. Astron. Soc.* **452,** 3869-3893 (2015).
48. Inserra, C. et al. A Statistical Approach to Identify Superluminous Supernovae and Probe Their Diversity. *Astrophys. J.* **854,** 175 (2018).
49. Quimby, R. et al. Spectra of Hydrogen-poor Superluminous Supernovae from the Palomar Transient Factory. *Astrophys. J.* **855,** 2 (2018).
50. Liu, J.-Q., Modjaz, M. & Bianco, F. Analyzing the Largest Spectroscopic Data Set of Hydrogen-poor Super-luminous Supernovae. *Astrophys. J.* **845,** 85-(2017).
51. Inserra, C. et al. Euclid: Superluminous supernovae in the Deep Survey. *Astron. Astrophys.* **609,** A83-(2018).
52. Inserra, C. et al. Complexity in the light curves and spectra of slow-evolving superluminous supernovae. *Mon. Not. R. Astron. Soc*. **468,** 4642-4662 (2017).

53. Nicholl, M. et al. SN 2015BN: A Detailed Multi-wavelength View of a Nearby Superluminous Supernova. *Astrophys. J.* **826,** 39 (2016).
54. Nicholl, M. et al. One Thousand Days of SN2015bn: HST Imaging Shows a Light Curve Flattening Consistent with Magnetar Predictions. *Astrophys. J.* **866,** L24 (2018).
55. Vreeswijk, P. et al. On the Early-time Excess Emission in Hydrogen-poor Superluminous Supernovae. *Astrophys. J.* **835,** 58 (2017).
56. Lunnan, R. et al. PS1-14bj: A Hydrogen-poor Superluminous Supernova With a Long Rise and Slow Decay. *Astrophys. J.* **831,** 144-(2016).
57. Leloudas, G. et al. SN 2006oz: rise of a super-luminous supernova observed by the SDSS-II SN Survey. *Astron. Astrophys.* **541,** A129-(2012).
58. Smith, M. et al. DES14X3taz: A Type I Superluminous Supernova Showing a Luminous, Rapidly Cooling Initial Pre-peak Bump. *Astrophys. J.* **818,** L8-(2016).
59. Nicholl, M. et al. LSQ14bdq: A Type Ic Super-luminous Supernova with a Double-peaked Light Curve. *Astrophys. J.* **807,** L18 (2015).
60. Nicholl, M. & Smartt, S. J. Seeing double: the frequency and detectability of double-peaked superluminous supernova light curves. *Mon. Not. R. Astron. Soc.* **457,** L79-L83 (2016).
61. Nicholl, M. et al. Superluminous supernovae from PESSTO. *Mon. Not. R. Astron. Soc.* **444,** 2096-2113 (2014).
62. Blanchard, P. K. et al. The Type I Superluminous Supernova PS16aqv: Lightcurve Complexity and Deep Limits on Radioactive Ejecta in a Fast Event. *Astrophys. J.* **865,** 9 (2018).
63. Chen, T.-W. et al. The host galaxy and late-time evolution of the superluminous supernova PTF12dam. *Mon. Not. R. Astron. Soc.* **452,** 1567-1586 (2015).
64. Kangas, T. et al. Gaia16apd - a link between fast and slowly declining type I superluminous supernovae. *Mon. Not. R. Astron. Soc.* **469,** 1246-1258 (2017).
65. Moriya, T., Sorokina, E. & Chevalier, R. A. Superluminous Supernovae. *SSRv* **214,** 59 (2018).
66. Chomiuk, L. et al. Pan-STARRS1 Discovery of Two Ultraluminous Supernovae at z~0.9. *Astrophys. J.* **743,** 114-(2011).
67. Berger, E. et al. Ultraluminous Supernovae as a New Probe of the Interstellar Medium in Distant Galaxies. *Astrophys. J.* **755** L29-(2012).
68. Howell, D. A. et al. Two Superluminous Supernovae from the Early Universe Discovered by the Supernova Legacy Survey. *Astrophys. J.* **779** 98-(2013).
69. Vreeswijk, P. et al. The Hydrogen-poor Superluminous Supernova iPTF 13ajg and its Host Galaxy in Absorption and Emission. *Astrophys. J.* **797,** 24 (2014).
70. Yan, L. et al. Far-ultraviolet to Near-infrared Spectroscopy of a Nearby Hydrogen-poor Superluminous Supernova Gaia16apd. *Astrophys. J.* **840,** 57-(2017).
71. Pan, Y.-C. et al. DES15E2mlf: a spectroscopically confirmed superluminous supernova that exploded 3.5 Gyr after the big bang. *Mon. Not. R. Astron. Soc.* **470,** 4241-4250 (2017).
72. Modjaz, M. et al. The Spectral SN-GRB Connection: Systematic Spectral Comparisons between Type Ic Supernovae and Broad-lined Type Ic Supernovae with and without Gamma-Ray Bursts. *Astrophys. J.* **832,** 108 (2016).
73. Yan, L. et al. Detection of Broad Hα; Emission Lines in the Late-time Spectra of a Hydrogen-poor Superluminous Supernova. *Astrophys. J.* **814,** 108 (2015).
74. Yan, L. et al. Hydrogen-poor Superluminous Supernovae with Late-time Hα Emission: Three Events From the Intermediate Palomar Transient Factory. *Astrophys. J.* **848,** 6 (2017).
75. Modjaz, M., Gutiérrez, C. P. & Arcavi I. New regimes in the observation of core-collapse supernovae. *Nat. Astron.* https://doi.org/10.1038/s41550-019-0856-2 (2019).
76. Chugai, N. N., Chevalier, R. A. & Utrobin, V. P. Optical Signatures of Circumstellar Interaction in Type IIP Supernovae. *Astrophys. J.* **662,** 1136-1147 (2007).
77. Gutiérrez, C. P. et al. Type II Supernova Spectral Diversity. I. Observations, Sample Characterization, and Spectral Line Evolution. *Astrophys. J.* **850,** 89-(2017).
78. Nicholl, M. et al. Nebular-phase Spectra of Superluminous Supernovae: Physical Insights from Observational and Statistical Properties. *Astrophys. J.* **871,** 102-(2019).

79. Jerkstrand, A. et al. Long-duration Superluminous Supernovae at Late Times. *Astrophys. J.* **835,** 13-(2017).
80. Levan, A. et al. Superluminous X-Rays from a Superluminous Supernova. *Astrophys. J.* **771,** 136 (2013).
81. Margutti, R. et al. Results from a Systematic Survey of X-Ray Emission from Hydrogen-poor Superluminous SNe. *Astrophys. J.* **864,** 45 (2018).
82. Bhirombhakdi, K. et al. Where is the Engine Hiding Its Missing Energy? Constraints from a Deep X-Ray Non-detection of the Superluminous SN 2015bn. *Astrophys. J.* **868,** L32-(2018).
83. Chen, T.-W. et al. The evolution of superluminous supernova LSQ14mo and its interacting host galaxy system. *Astron. Astrophys*. **602,** A9-(2017).
84. Bose, S. et al. Gaia17biu/SN 2017egm in NGC 3191: The Closest Hydrogen-poor Superluminous Supernova to Date Is in a Normal, Massive, Metal-rich Spiral Galaxy. *Astrophys. J.* **853,** 57-(2018).
85. Coppejans, D. L. et al. Jets in Hydrogen-poor Superluminous Supernovae: Constraints from a Comprehensive Analysis of Radio Observations. *Astrophys. J.* **856,** 56-(2018).
86. Eftekhari, T, et al. A Radio Source Coincident with the Superluminous Supernova PTF10hgi: Evidence for a Central Engine and an Analog of the Repeating FRB 121102?. *Astrophys. J.* **876,** L10-(2019).
87. Inserra, C., Bulla, M., Sim, S. A., Smartt, S. J. Spectropolarimetry of Superluminous Supernovae: Insight into Their Geometry. *Astrophys. J.* **831,** 79-(2016).
88. Leloudas, G. et al. Time-resolved Polarimetry of the Superluminous SN 2015bn with the Nordic Optical Telescope. *Astrophys. J.* **837,** L14-(2017).
89. Leloudas, G. et al. Polarimetry of the Superluminous Supernova LSQ14mo: No Evidence for Significant Deviations from Spherical Symmetry. *Astrophys. J.* **815,** L10-(2015).
90. Lunnan, R. et al. A UV resonance line echo from a shell around a hydrogen-poor superluminous supernova. *NatAs* **2,** 887-895 (2018).
91. Tanaka, M. et al. Rapidly Rising Transients from the Subaru Hyper Suprime-Cam Transient Survey. *Astrophys. J.* **819,** 5-(2016).
92. Kasliwal, M. et al. Rapidly Decaying Supernova 2010X: A Candidate ".Ia" Explosion. *Astrophys. J.* **723,** L98-L102 (2010).
93. Drout, M. et al. The Fast and Furious Decay of the Peculiar Type Ic Supernova 2005ek. *Astrophys. J.* **774,** 58-(2013).
94. Moriya, T. et al. Light-curve and spectral properties of ultrastripped core-collapse supernovae leading to binary neutron stars. *Mon. Not. R. Astron. Soc.* **466,** 2085 (2017).
95. Perets, H. et al. A faint type of supernova from a white dwarf with a helium-rich companion. *Nature* **465,** 322-325 (2010).
96. Shen, K. et al. Thermonuclear .Ia Supernovae from Helium Shell Detonations: Explosion Models and Observables. *Astrophys. J.* **715,** 767-774 (2010).
97. Inserra, C. et al. OGLE-2013-SN-079: A Lonely Supernova Consistent with a Helium Shell Detonation. *Astrophys. J.* **799,** L2-(2015).
98. Pastorello, A. et al. Massive stars exploding in a He-rich circumstellar medium - IX. SN 2014av, and characterization of Type Ibn SNe. *Mon. Not. R. Astron. Soc.* **456,** 853-869 (2016).
99. Hosseinzadeh, G. et al. Type Ibn Supernovae Show Photometric Homogeneity and Spectral Diversity at Maximum Light. *Astrophys. J.* **836,** 158-(2017).
100. Jha, S. W., Maguire, K. & Sullivan, M. Observational properties of thermonuclear supernovae. *Nat Astron* https://doi.org/10.1038/s41550- 019-0858-0 (2019).
101. Prentice, S. et al. The Cow: Discovery of a Luminous, Hot, and Rapidly Evolving Transient. *Astrophys. J.* **865,** L3-(2018).
102. Perley, D. et al. The fast, luminous ultraviolet transient AT2018cow: extreme supernova, or disruption of a star by an intermediate-mass black hole?. *Mon. Not. R. Astron. Soc.* **484,** 1031-1049 (2019).

103. Margutti, R. et al. An Embedded X-Ray Source Shines through the Aspherical AT2018cow: Revealing the Inner Workings of the Most Luminous Fast-evolving Optical Transients. *Astrophys. J.* **872,** 18 (2019).
104. Kuin, N. P. M. et al. Swift spectra of AT2018cow: A White Dwarf Tidal Disruption Event?. *Mon. Not. R. Astron. Soc*. **487,** 2505-2521 (2019).
105. Fox, O. & Smith, N. Signatures of Circumstellar Interaction in the Unusual Transient AT2018cow. Preprint at arXiv:1903.01535-(2019).
106. Arcavi, I. et al. Rapidly Rising Transients in the Supernova - Superluminous Supernova Gap. *Astrophys. J*. **819,** 35-(2016).
107. Rest, A. et al. A fast-evolving luminous transient discovered by K2/Kepler. *NatAs* **2,** 307-311 (2018).
108. Vinko, J. et al. A Luminous, Fast Rising UV-transient Discovered by ROTSE: A Tidal Disruption Event? *Astrophys. J*. **798,** 12 (2015).
109. Ho, A. Y. Q. et al. The Death Throes of a Stripped Massive Star: An Eruptive Mass-Loss History Encoded in Pre-Explosion Emission, a Rapidly Rising Luminous Transient, and a Broad-Lined Ic Supernova SN2018gep. Preprint at arXiv:1904.11009-(2019).
110. Chen, P. et al. The Most Rapidly-Declining Type I Supernova 2019bkc/ATLAS19dqr. Preprint at arXiv:1905.02205 (2019).
111. Smith, N. et al. Coronal Lines and Dust Formation in SN 2005ip: Not the Brightest, but the Hottest Type IIn Supernova. *Astrophys. J*. **695,** 1334-1350 (2009).
112. Smith, N. et al. SN 2011hw: helium-rich circumstellar gas and the luminous blue variable to Wolf-Rayet transition in supernova progenitors. *Mon. Not. R. Astron. Soc.* **426,** 1905-1915 (2012).
113. Pastorello, A. et al. Massive stars exploding in a He-rich circumstellar medium - IV. Transitional Type Ibn supernovae. *Mon. Not. R. Astron. Soc.* **449,** 1921-1940 (2015).
114. Rivera Sandoval, L. E. et al. X-ray Swift observations of SN 2018cow. *Mon. Not. R. Astron. Soc*. **480,** L146-L150 (2018).
115. Poznanski, D. et al. An Unusually Fast-Evolving Supernova. *Science* **327,** 58 (2010).

**Acknowledgements**
The author would like to thank D. Perley, S. Prentice and M. Pursiainen for sharing their dataset on Fast Blue, Optical Transients. I would also like to thank G. Leloudas and an anonymous referee to have improved the overall manuscript.

**Competing interests**
The authors declare no competing interests.

**Additional information**